\documentstyle[12pt,aaspp4]{article}
\begin{document}
\title{Arm Structure in Anemic Spiral Galaxies}
\author{Debra Meloy Elmegreen\altaffilmark{1},
Bruce G. Elmegreen\altaffilmark{2},
Jay A. Frogel\altaffilmark{3,4},
Paul B. Eskridge\altaffilmark{5}, Richard W.
Pogge\altaffilmark{3},
Andrew Gallagher\altaffilmark{1},
and Joel Iams\altaffilmark{1,6}}
\altaffiltext{1}{Department of Physics and Astronomy,
Vassar College,
Poughkeepsie, NY 12604;
e--mail: elmegreen@vassar.edu, angallagher@vassar.edu}
\altaffiltext{2}{IBM T. J. Watson Research Center,
P.O. Box 218, Yorktown
Heights, NY 10598, e--mail:bge@watson.ibm.com}
\altaffiltext{3}{Dept. of Astronomy, Ohio State
University, Columbus, OH,
e--mail:pogge@astronomy.ohio-state.edu }
\altaffiltext{4}{Code 52, NASA Headquarters,
Washington, D.C., e--mail jfrogel@hq.nasa.gov}
\altaffiltext{5}{Dept. of Astronomy, Minnesota State
University, Mankato, MN
56001, e--mail:paul.eskridge@mnsu.edu}
\altaffiltext{6}{Dept. of Astronomy, Williams College,
Williamstown, MA,
e--mail: jiams1@williams.edu}
\begin{abstract}
Anemic galaxies have less prominent star formation
than normal
galaxies of the same Hubble type. Previous studies
showed they are
deficient in total atomic hydrogen but not in
molecular hydrogen.
Here we compare the combined surface densities of HI
and H$_2$ at
mid-disk radii with the Kennicutt threshold for star
formation.
The anemic galaxies are below threshold, which
explains their lack
of prominent star formation, but they are not much
different than
other early type galaxies, which also tend to be below
threshold.
The spiral wave amplitudes of anemic and normal
galaxies were also
compared, using images in B and J passbands from the
OSU Bright Spiral Galaxy Survey.  Anemic galaxies have
normal
spiral wave properties too, with the same amplitudes
and radial
dependencies as other galaxies of the same arm class.
Because of the lack of gas,
spiral waves in early type galaxies and anemics do not
have a
continuous supply of stars with low velocity
dispersions
to maintain a marginally stable
disk. As a result, they are either short-lived,
evolving toward
lenticulars and S0 types in only a few rotations at
mid-disk, or
they are driven by the asymmetries associated with gas
removal in
the cluster environment.
\end{abstract}

\section{Introduction}

Galaxies with smooth spiral arms in their main disks
on the Palomar Sky
Survey were
classified as anemic by van den Bergh (1976). He
described them as
spirals that are gas-poor with characteristics
intermediate
between S0's and normal spirals.

Anemics are most often in rich clusters, but they are
also in the
field.  The field population of anemics may not be
different than
normal galaxies because nearly all Sa galaxies, even
those in the
field, were classified as anemic by van den Bergh
(1976) as a
result of their generally indistinct spiral arms
(Rubin et al.
1978; Bothun \& Sullivan 1980).

Anemics in clusters seem to result from gas stripping
(van den
Bergh 1976). Not all clusters have HI deficiencies
(Bothun 1982;
Giovanelli \& Haynes 1985), and even when they do
there is not a
good correlation between HI deficiency, red color, and
anemic type
(Kennicutt, Bothun, \& Schommer 1984). In Virgo, where
gas
stripping is clear (Chamaraux, Balkowski, \& G\'erard
1980;
Cayatte et al. 1994), the star formation rates and
colors of
stripped galaxies are about normal for their relative
HI contents,
and all three properties are similar to those of field
galaxies
with earlier Hubble types (Kennicutt 1983).  This
implies that
stripping is a gradual process for a galaxy as a
whole, mimicking
in some respects normal galaxy evolution (Kennicutt
1983).
Stripped galaxies that are anemics may be somewhat
special, having
radial orbits in the cluster (Vollmer et al. 2001a).

Gas stripping in clusters has a much more obvious
signature in the
distribution of the HI gas. Stripped galaxies in the
centers of
clusters have relatively small radii for their HI
disks compared
to their optical disk (Giovanelli \& Haynes 1983; van
Gorkom \&
Kotanyi 1985; Warmels 1988; Cayatte et al. 1994;
Bravo-Alfaro et
al. 2000), and where the HI remains, the HI surface
density can be
relatively low too (Kenney \& Young 1989).  Anemics
have the
largest HI anomalies, with the smallest relative HI
radii and the
largest inner disk HI deficiencies (Cayatte et al.
1994;
Bravo-Alfaro et al. 2001).

Gas stripping in HI hardly affects the molecular mass
and surface
density, however (Kenney \& Young 1989; Casoli et al.
1991;
Boselli et al. 1997; Vollmer et al. 2001b). The H$_2$
mass may
even be enhanced relative to HI in the inner regions,
as if the
increased pressure associated with stripping produces
denser
clouds (Kenney \& Young 1989; Cayatte et al. 1994). 
This
stripping pressure can also increase the star
formation rates
along the leading edges of galaxies if stripped gas
falls back
(Vollmer et al. 2001b), and it can make the inner CO
disk
asymmetric (Kenney et al. 1990; Hidaka \& Sofue 2002).

Here we examine the surface densities at mid-radii for
the total atomic
and molecular gas in Virgo galaxies and other galaxies
and compare them
with the surface density threshold for star formation
found by Kennicutt
(1989).   We find that the total surface densities are
below the threshold
in the main disks of anemics, as is often the case for
early type galaxies
(Caldwell et
al. 1991), but they exceed the threshold in later type
galaxies. This
threshold difference at mid-disk radius explains the
low star formation
activity in anemic galaxies.

The spiral arm strength in anemics is examined next
using blue and
near-infrared images from the OSU Bright Spiral Galaxy
Survey. We find that anemics have normal arm/interarm
contrasts for their spiral arm classes (grand design
versus flocculent).
The presence of normal stellar density waves in
galaxies with low gas
densities is unusual because gas is a strong amplifier
for stellar wave
instabilities (Bertin \& Romeo 1988).  Either the
stripping process
and the associated gas asymmetries (Kenney et al.
1990) are driving the
stellar spirals, or the stripping took place very
recently.

\section{Gas Surface Densities in Anemic and Normal
Galaxies}

Anemic galaxies have normal total CO masses for their
Hubble types
(Kenney \& Young 1989).  Previous studies did not
compare the
H$_2$ surface densities in anemics and normal
galaxies, nor did
they compare the total atomic and molecular surface
densities in
anemic galaxies to the Kennicutt (1989) threshold for
star
formation.

The bottom panel in Figure 1 shows all of the detected
H$_2$
surface densities, $\Sigma$, in the study by Rownds \&
Young (1999).
Multiple
detections in the same galaxy are plotted as multiple
points, and
non-detections are plotted at the bottom of the panel,
where
$\log\Sigma=-1$. The anemic nature of the galaxies was
determined
from the classifications in van den Bergh (1976) and
van den
Bergh, Pierce, \& Tully (1990).  Anemic galaxies are
plotted with
open circles in the figure (NGC 3718, NGC 4293, NGC
4450, NGC
4522, NGC 4548, NGC 4569, NGC 4579, NGC 4651, NGC
4689, NGC 4710,
NGC 4826), and normal galaxies are plotted with dots.
The abscissa
is the morphological $T$ type in the Third Reference
Catalogue of
Galaxies (hereafter RC3; de Vaucouleurs, et al. 1991);
$T=1$ is
approximately equivalent to Hubble type Sa, $T=8$ is
equivalent to
Sd. The conversion factor from CO to H$_2$ varies a
little with
Hubble type (Boselli, Lequeux, \& Gavazzi 2002), as do
fractional
gas masses and other gas properties, so we plot all
our results
as a function of $T$ type to remove these effects from
our discussion.  The
figures indicate that there is no significant
difference in the
point-by-point distribution of H$_2$ surface density
for anemics
and normal galaxies. This is consistent with the lack
of any such
difference in the total H$_2$ mass and the average
H$_2$ surface
density (e.g., Kenney \& Young 1989).

The top panel of Figure 1 shows the average central HI
surface
density versus the $T$ type for all of the galaxies in
Warmels
(1988) and Cayatte et al. (1994).  This average is the
integral of
the radius times the HI surface density out to half of
the optical
radius ($D_0/4$ for corrected optical diameter $D_0$
in the RC3),
divided by the integral of the radius out to this same
distance
(as tabulated by Cayatte et al. for their galaxies and
calculated
here from data in Warmels). The average surface
density
inside half the optical
radius is usually representative of the inner disk
surface density
because the intensity of HI is relatively constant
there. There is
a well-known correlation between average HI surface
density and
morphological type, with later types having higher HI
surface
densities, and there is another well-known correlation
with
anemics having lower mid-disk surface densities than
normal
galaxies within a $T$ type (Kenney \& Young 1989;
Cayatte et al.
1994).  The anemics in this figure are NGC 4450, NGC
4548, NGC 4569, NGC
4579,
NGC 4651, and NGC 4689.

The sum of the HI and H$_2$ surface densities
determines the star
formation properties in a galaxy. Kennicutt (1989) and
Martin \&
Kennicutt (2001) found that stars tend to form where
the summed
surface density, $\Sigma_{tot}$, exceeds about 0.7
times the Toomre
(1964) critical surface density,
$\Sigma_{crit}$=$\kappa$c/(3.36G),
for an assumed velocity dispersion of c=6 km s$^{-1}$
and an
epicyclic frequency $\kappa$ derived from the rotation
curve. To
determine $\Sigma_{tot}$ for this study, we used the
average HI
surface density inside $D_0/4$, as plotted in Figure
1, and the
H$_2$ surface density at the radius $D_0/4$ from
Rownds \& Young
(1999). We determined $\kappa$ at $D_0/4$ from the HI
rotation
curves in Guhathakurta et al. (1988).  There are 5
anemics in this
overlapping sample, NGC 4548, NGC 4550, NGC 4569, NGC
4579, and
NGC 4689.

Figure 2 suggests that anemics are not significantly
different
from other early-type galaxies, which all have a low
ratio
$\Sigma$/$\Sigma_{crit}$, as found by Caldwell et al.
(1991). The
low value of this ratio for anemics and other early
types is the
result of a moderately low H$_2$ surface density
compared to that
in later type spirals, a very low HI surface density,
and a high
$\kappa$ in the early types, which comes from the
relatively massive
bulge.

Anemics have unusually weak star formation in their
main disks, like other
very early
type spiral galaxies, because gravitational
instabilities and
other processes that normally promote star formation
are not
possible in these galaxies.  This is true even though
their total
molecular masses are normal for their Hubble types. 
Part of the
reason for the normal molecular masses in anemics is
their higher
molecular fractions at mid-disk radii (Kenney \& Young
1989;
Casoli et al. 1991).  The motion of the anemics
through the
intergalactic medium causes higher disk pressures even
in parts of
the disk where it does not strip the gas away.  
Higher pressures
are generally associated with greater molecular
fractions as a
result of increased cloud densities and self-shielding
(e.g.,
Hidaka \& Sofue 2002). Most of the extra molecular
material is
probably in the form of diffuse molecular clouds,
which do not
form stars. 

Some anemics have active star formation in their
inner regions, typically within 0.2R$_{25}$.  For
example, NGC 4548, a
strongly barred anemic galaxy, has star formation
knots in the spiral arms
at the ends of the bar.  NGC 4580 and NGC 4689 have
circumnuclear rings of
star formation and smooth arms elsewhere. We have not
checked if the 
gas surface density exceeds the critical threshold in
these inner regions.
Our results apply only to the mid-disk positions,
where HI and
CO data are available.

Guiderdoni (1987) originally suggested that surface
density
might provide a threshold for star formation, based on
the
HI surface densities in anemics.  He considered an
absolute threshold rather than a dynamical one that
varies with
$\kappa$, and did not include H$_2$, but he was
essentially
correct in his conclusions about star formation.

\section{Spiral Arm Strengths in Anemic and Normal
Galaxies}
\subsection{Observations}

Anemic galaxies have spiral arms, so the lack of star
formation
could in principle be related also to a low spiral arm
strength.
To check this, we measured arm-interarm contrasts in B
and J bands
for all the galaxies in Virgo that were observed as
part of the
Ohio State University Bright Spiral Galaxy Survey
(Frogel,
Quillen, \& Pogge 1996; Eskridge et al. 2002). For
those galaxies
in which J-band data had poor flat fields or were
unavailable, K
or H band data were substituted. The images were
obtained with the
1.8-m telescope at Lowell Observatory and the 1.5-m
telescope at
Cerro Tololo Inter-American Observatory, using the
Ohio State
Infrared Imager/ Spectrometer (OSIRIS) and the Cerro
Tololo
Infrared Imager (CIRIM). The plate scale is between
1.05 arcsec
and 1.55 arcsec per pixel in J band for the different
instrument/telescope combinations. NIR passbands are
particularly
useful for highlighting the underlying old stellar
populations,
with average ages in excess of 10$^{10}$ yrs (see
review in
Frogel, Quillen, \& Pogge 1996), whereas blue light is
dominated
by contributions from younger stellar populations.

Our sample consists of 30 galaxies: 12 anemics
(classified by van
den Bergh 1976) and 18 normal spirals, selected to
have a wide
variety of Hubble types, arm classes, and gas content.
The OSU
survey is magnitude-limited, and all anemic galaxies
in that
survey are included here. All galaxies with types S0
through Sb in
our sample are barred or ovally distorted, while later
types
include nonbarred galaxies. The properties of the
observed
galaxies are listed in Tables 1 (anemics) and 2
(normal galaxies)
with their NGC designation, morphological $T$ type,
optical arm class (from Elmegreen \& Elmegreen
1987, where F is flocculent, G is grand design, and M
is multiple
arm), radius R$_{25}$ (where the surface brightness
drops to 25
mag arcsec$^{-2}$) in arcsec, arm-interarm magnitude
contrast
$\Delta$m in B and J bands, described below, and HI
index (from
RC3).  High values of the HI index correspond to an HI
deficiency.

The images were flat-fielded and sky-subtracted
following
procedures described by Berlind et al. (1997).
Combined images
were deprojected by using the position angles and
inclinations
listed in the RC3. Because this study considers only
relative
magnitudes, absolute calibrations were not necessary.
A polar plot
of (r,$\theta$), for distance r from the galaxy center
and
azimuthal angle $\theta$ around the galaxy, was made
of each
galaxy using a script written in IRAF.  Azimuthal
intensity cuts
(corresponding to horizontal cuts on the (r,$\theta$)
plots) 3
pixels wide in radius were taken using the IRAF task
PVECTOR.
These cuts were repeated for several different radial
distances
from the center out to the optical edge of the galaxy
at $\sim$
R$_{25}$, in steps of 10 pixels in radius
(approximately every
0.05 R$_{25}$). From these profiles, the arm and
interarm
intensities and their magnitude differences
(arm-interarm
contrasts) were measured. The results for each
anemic galaxy were then compared to the results of
normal galaxies
of the same or similar type, considering also their
arm class and
HI index.

\subsection{Arm-Interarm Contrasts}

As described in the previous section, the
arm-interarm contrasts were measured for this sample
of galaxies
at regularly spaced radii throughout the disk. These
contrasts
were found to have the usual properties, namely, they
are larger
in grand design galaxies than in flocculent galaxies,
they
increase with radius for non-barred galaxies and
decrease with
radius outside the bar for barred galaxies, and the
arms are bluer
in flocculent galaxies than in grand design galaxies.
In addition, the blue
arm-interarm
contrasts fluctuate with radius much more in normal
galaxies than
in anemics, which is an obvious result of the smoother
arms in
anemics. The radial fluctuations of blue arm-interarm
contrast for
grand design normal galaxies is $\pm0.7$ mag, while in
grand
design anemic galaxies it is $\pm0.1$ mag.

The average arm-interarm B-band and J-band magnitude
contrasts are
shown as a function of galaxy $T$ type in Figure 3,
separated into
different panels by color (top is blue and bottom is
J) and arm
class (left is grand design or multiple arm, right is
flocculent).
Normal galaxies have solid symbols and anemic galaxies
have open
symbols.  The grand design and multiple arm galaxies
of all Hubble
types, whether anemic or normal, have higher average
arm-interarm
contrasts than the flocculent galaxies. For either arm
class,
there is no distinction in arm-interarm contrast
between anemic
and normal galaxies.  Rubin et al. (1978) also noted
that the
arm-interarm contrast in an anemic galaxy looked
normal.

This result implies that density waves in anemic
galaxies are as
strong as density waves in normal galaxies.  Anemic
structure is
therefore the result of a low
$\Sigma_{tot}/\Sigma_{crit}$.

\section{Discussion and Conclusions}

Anemic galaxies are spiral galaxies of normal size
that have lost gas over
time from a combination of internal star formation and
tidal stripping.
They have normal spiral density wave properties, but
those in clusters
have truncated outer HI disks and low total gas
surface densities.
Their main disk spiral arms are smoother than usual
because of a lack of
giant
HII regions and star complexes, but they have
approximately normal star
formation rates and colors for their gas content,
i.e., normal star
formation efficiencies (as determined from data in
Rownds \& Young 1999).

The lack of star formation in anemics seems to be the
result of their
low $\Sigma/\Sigma_{crit}$.  This is the first check
on the utility of
this star formation criterion that does not rely on a
comparison between
galaxies of intrinsically different types or between
star formation
regions with very different radii in a galaxy. For
example, other
studies of $\Sigma_{tot}/\Sigma_{crit}$ emphasized the
outer parts of
late type spirals (Kennicutt 1989; Martin \& Kennicutt
2001) or nuclear
starburst regions (Shlosman \& Begelman 1989;
Elmegreen 1994), or they
were for elliptical galaxies (Vader \& Vigroux 1991)
or low surface
brightness galaxies (van der Hulst et al. 1993). Here
we see the effects
of decreasing $\Sigma/\Sigma_{crit}$ over time
throughout the whole
disk in an otherwise normal galaxy.  When
$\Sigma/\Sigma_{crit}$ drops
below 0.7 as a result of gas stripping or other gas
removal processes,
the star formation rate per unit area drops and the
giant HII regions
and OB associations disappear.

The future of anemic galaxies may be understood from
numerical models
of disk galaxies that have no source of cooling for
the stellar population
and no
continuous supply of fresh stars with low velocity
dispersions.  In these
models (Sellwood \& Carlberg 1984; Fuchs \& von Linden
1998), the stellar
disk rapidly heats up and the spiral waves stop.
Anemics are
therefore evolving toward lenticulars or S0's as a
result of their
low gas abundance. 
The observation of strong spiral waves
in severely stripped galaxies also implies three other
things:
(1) stellar spirals need stars with low velocity
dispersions
and cool gas mostly in their inner regions,
which are still not heavily depleted in anemics; (2)
the stripping probably occurred within the last few
orbit times at mid-disk
radius because spiral wave amplification gets weak
quickly when both
the stellar and gaseous disks are Toomre-stable (Jog
\& Solomon 1984;
Bertin \& Romeo 1988), and (3) the stripping process
may
itself drive spiral structure, perhaps through the
asymmetric mass
distribution of the stripped gas. Such asymmetries are
present in both
the HI (Boselli et al. 1994; Bravo-Alfaro et al. 1997,
2000) and CO
(Kenney et al. 1990).

Acknowledgements:
We thank the many OSU graduate students who collected
data for
this project, and wish to \ especially note the many
nights of
work that Ray Bertram and Mark Wagner have devoted to
the OSU
survey. The OSU Survey project was generously
supported by grants AST-9217716
and AST-9617006 from the National Science Foundation
to OSU (JAF as PI). We thank OSU
graduate students Glenn Tiede and Leslie Kuchinski and
Vassar
undergraduate Theresa Brandt for assistance with the
data
reduction. We acknowledge support from the Vassar URSI
program and
the Wm. F. Keck Foundation for financing the
undergraduates at
Vassar. This research has made use of the NASA/IPAC
Extragalactic
Database (NED) which is operated by the Jet Propulsion
Laboratory,
California Institute of Technology, under contract
with the
National Aeronautics and Space Administration, and
IRAF, distributed by the
National Optical Astronomy Observatories, which are
operated by the
Association of Universities for Research in Astronomy,
Inc., under
cooperative agreement with the National Science
Foundation.

\begin{table}
\label{table:anemic}
\caption{Anemic Galaxies}
\begin{center}
\begin{tabular}{ccccccc}
Name&Type&Arm Class&R$_{25}$&$\Delta$ m$_{B}$&$\Delta$
m$_{J}$&HI\\
1302    & RSBR0&  F&     116.7&0.70&   0.43 &---\\
4314    & SBT1&  G  & 125.8&2.10& 2.00&  1.2\\
4457    & RSXS0&  G&     80.8   &--- &0.64&---\\
4548& SBT3&  G   &161.1&1.33&   1.22  & 3.91\\
4580    & SXT1&  M/F&    62.7& 0.87&  0.57&   4.91\\
4643    & SBT0&  G   &92.7&1.80 &  1.34 &---\\
4651    & SAT5&  M/F&    119.4& 0.80 & 0.36&   1.86\\
4689    & SAT4&  F  & 128.0&  0.34& 0.43 &  3.22\\
4691& RSBS0&  G&     84.6   &1.73& 1.36&---\\
4699 &SXT3  & F&     114.1&0.48&  0.25 &   3.08\\
4941    & RSXR2&  M& 108.9& 0.55&  0.26 &  3.45\\
5101     &RSBT0&  G&     161.1&1.46&   1.13 &---\\
\end{tabular}
\end{center}
\end{table}

\begin{table}
\label{table:normal}
\caption{Normal Galaxies}
\begin{center}
\begin{tabular}{ccccccc}
Name&Type&Arm Class&R$_{25}$&$\Delta$ m$_{B}$&$\Delta$
m$_{J}$&HI\\
NGC 0157    &SXT4   & G &125.1&1.20&  0.76&   2.11\\
1792&   SAT4    & M &157.4&1.00&  0.52&2.9\\
3684&   SAT4    & F&    92.7    &1.20&  0.40 &  
1.41\\
3810    &SAT5   & F &123.0&0.89&      0.50 &   2.17\\
4051    &SXT4   & G &157.4&1.40&      0.85 &   2.41\\
4145&   SXT7     &M &176.7&0.65&      0.45  &  1.29\\
4212    &SA5     &M&    94.86    &1.30& 0.67    &---\\
4254    &SAS5   &M& 161.1   &1.02&  0.80   & 2.26\\
4303    &SXT4    &M&    194.0   &1.20&  0.94  & 2.14\\
4394&   RSBR3    &G&    108.9    &---& 1.12   &3.72\\
4414    &SAT5    &F&    108.9   &0.75&  0.40   &
2.04\\
4571    &SAR7 &  M& 108.9   &0.68&  0.33 &  2.90\\
4665    &SBS0   & G &114.1  &1.3 & 1.00&    ---\\
4775    &SAS7    &F&    64.14    &1.31& 0.65   &
2.03\\
5248    &SXT4    &G &185.0&1.47 &  0.70 &   1.67\\
5676    &SAT4    &M &119.4&0.75&      0.45  &  2.19 \\
IC4444&SXT4 & M &52.1   &1.5&  0.60  &  2.50\\
IC5325& SXT4    & M &82.6    &0.70& 0.36 &  3.03\\

\end{tabular}
\end{center}
\end{table}

\newpage
\begin{figure}
\vspace{6.5in}
\includegraphics{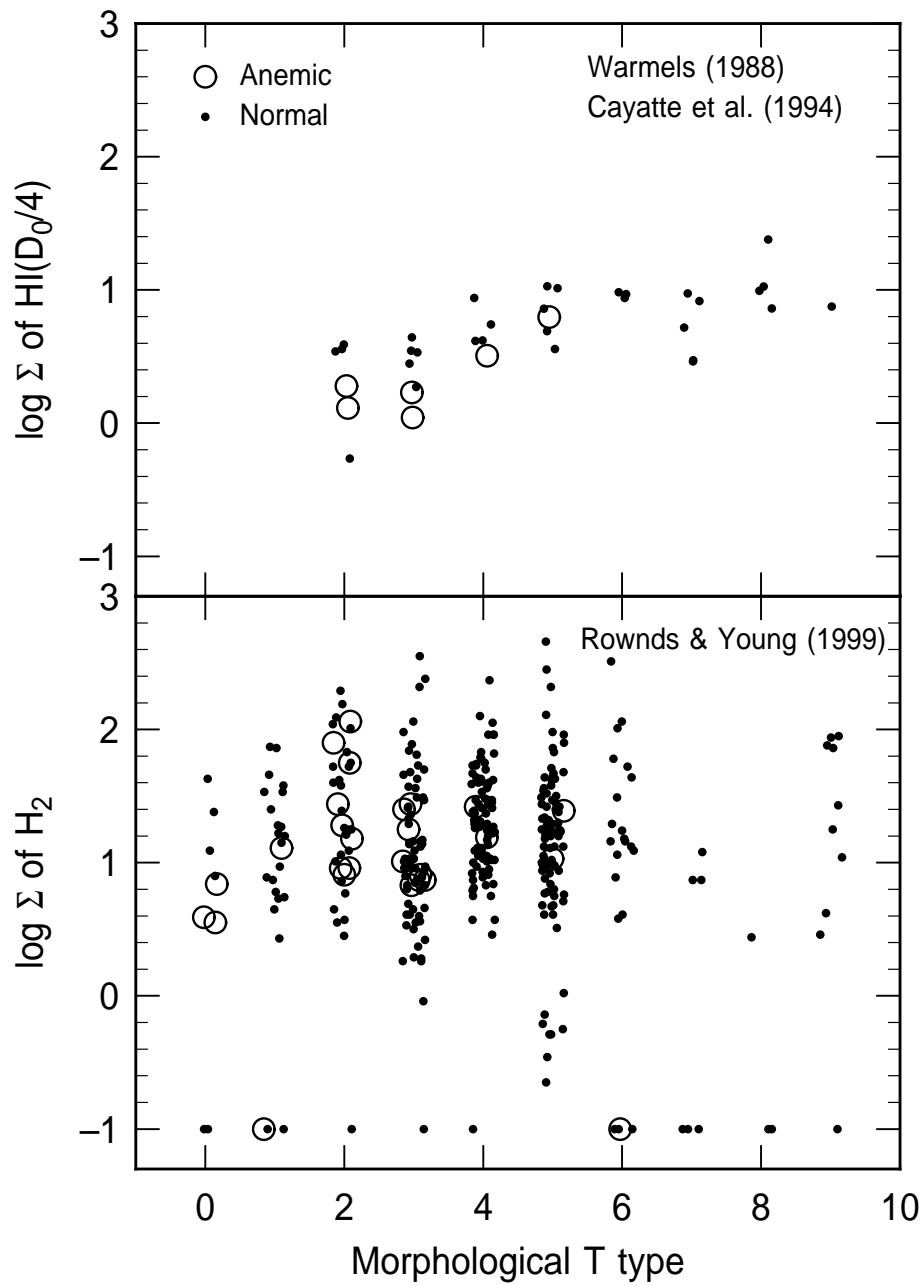}
\caption{The surface densities in M$_\odot$ pc$^{-2}$
of the
molecular (bottom panel) and atomic (top) phases in
galaxies are
plotted versus the morphological $T$ type (1=Sa,
2=Sab, etc.),
with early type galaxies on the left and late type
galaxies
on the right.  Open circles represent anemic galaxies
and dots
normal galaxies. The points in the bottom panel
include several
different radii in each of 121 different galaxies;
there are an
average of $\sim4$ points per galaxy, all from the
study by Rownds
\& Young (1999).  The points in the top panel are the
average HI
surface densities inside the half-light radii, from
Warmels (1988)
and Cayatte et al. (1994).  The anemic galaxies have
low HI
average surface densities but normal H$_2$ surface
densities for
their $T$ types, as noted in previous studies.}
\end{figure}

\newpage
\begin{figure}
\vspace{6.in}
\includegraphics{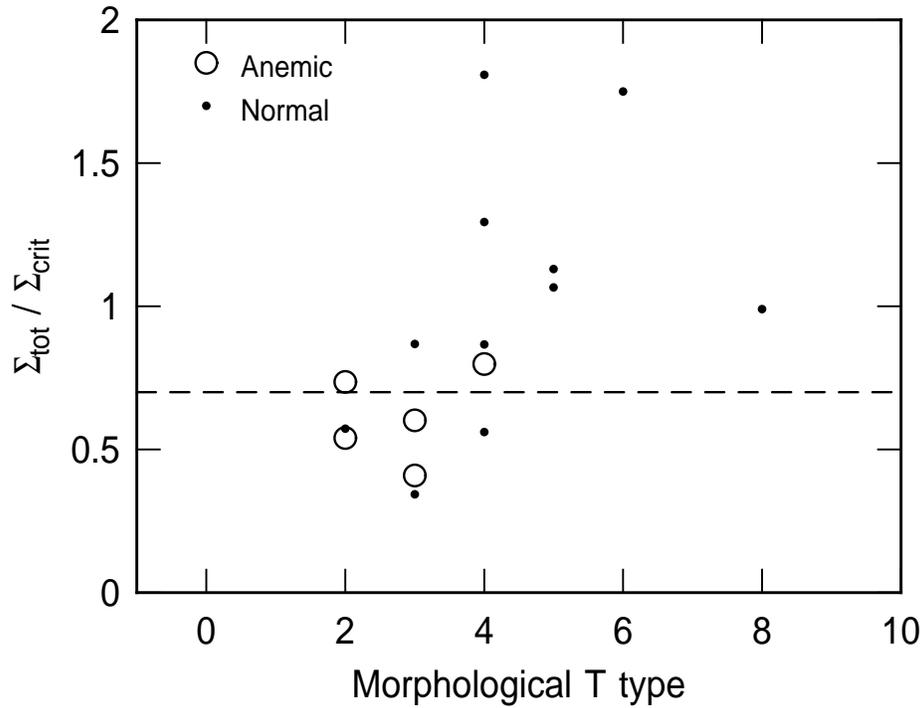}
\caption{The ratio of the summed molecular and atomic
surface
densities from Figure 1 to the critical value,
$\Sigma_{crit}$, is
plotted versus the morphological type for all galaxies
common to
the studies by Warmel (1988), Cayatte et al. (1994),
Rownds \&
Young (1999) and Guhathakurta et al. (1988).  All
quantities are
evaluated at about half the optical radius.
Anemic galaxies have surface densities below the
threshold for
star formation. }
\end{figure}

\newpage
\begin{figure}
\vspace{6.in}
\includegraphics{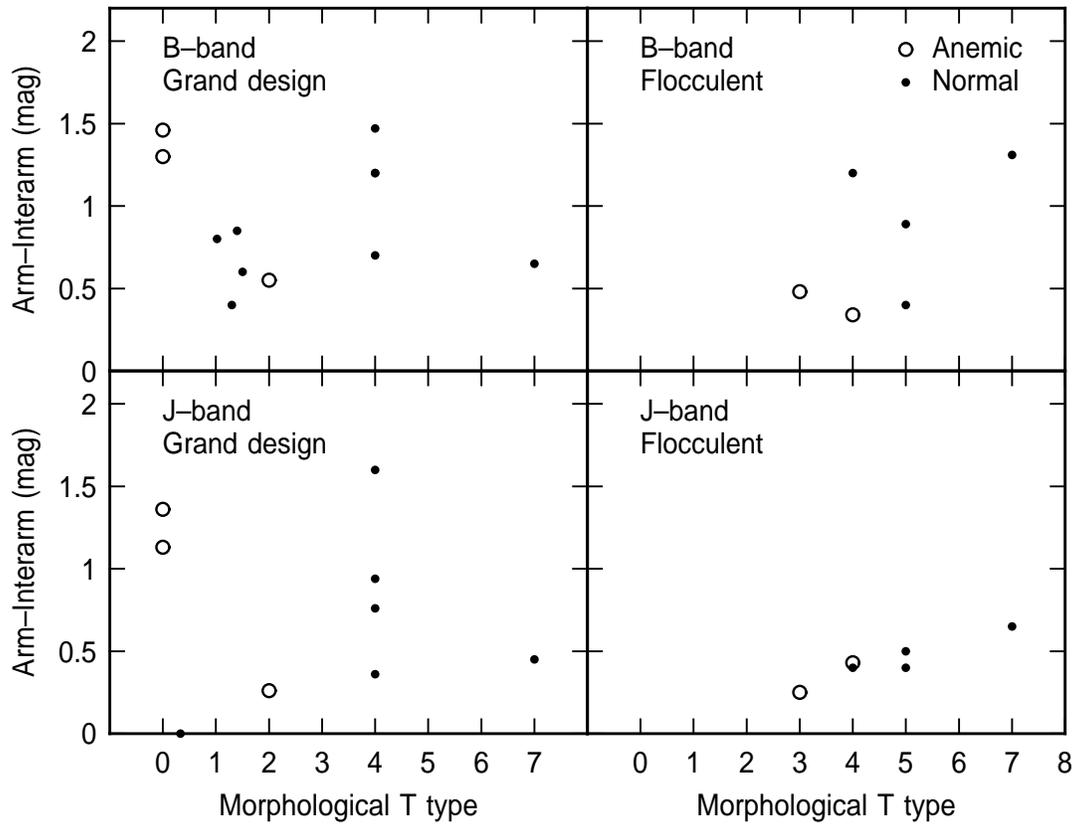}
\caption{Arm-interarm magnitude contrast versus
morphological
$T$ type for grand design and multiple arm galaxies on
the left and
flocculent galaxies on the right, in B passband at the
top and J
passband at the bottom.  Anemic galaxies have spiral
arm
properties that are indistinguishable from normal
galaxies.}
\end{figure}

\end{document}